\documentclass[review]{elsarticle}

\usepackage{lineno, hyperref, graphicx, amsmath, changepage, multirow, lscape}

\journal{International Journal of Human-Computer Studies}

\biboptions{authoryear}

\makeatletter
\def\@author#1{\g@addto@macro\elsauthors{\normalsize%
    \def\baselinestretch{1}%
    \upshape\authorsep#1\unskip\textsuperscript{%
      \ifx\@fnmark\@empty\else\unskip\sep\@fnmark\let\sep=,\fi
      \ifx\@corref\@empty\else\unskip\sep\@corref\let\sep=,\fi
      }%
    \def\authorsep{\unskip,\space}%
    \global\let\@fnmark\@empty
    \global\let\@corref\@empty  
    \global\let\sep\@empty}%
    \@eadauthor={#1}
}
\makeatother

\begin{document}

\begin{frontmatter}
\renewcommand*{\today}{February 8, 2018}
\title{Affective computing using speech and eye gaze: a review and bimodal system proposal for continuous affect prediction\tnoteref{mytitlenote}}
\tnotetext[mytitlenote]{Paper submission for review}

\author{Jonny O' Dwyer\corref{cor1}}
\ead{j.odwyer@research.ait.ie}
\cortext[cor1]{Corresponding author}

\author{Niall Murray}
\ead{nmurray@research.ait.ie}

\author{Ronan Flynn}
\ead{rflynn@ait.ie}

\address{Department of Electronics and Informatics,}
\address {Athlone Institute of Technology,}
\address {Athlone, Ireland}

\begin{abstract}
Speech has been a widely used modality in the field of affective computing. Recently however, there has been a growing interest in the use of multi-modal affective computing systems. These multi-modal systems incorporate both verbal and non-verbal features for affective computing tasks. Such multi-modal affective computing systems are advantageous for emotion assessment of individuals in audio-video communication environments such as teleconferencing, healthcare, and education. From a review of the literature, the use of eye gaze features extracted from video is a modality that has remained largely unexploited for continuous affect prediction. This work presents a review of the literature within the emotion classification and continuous affect prediction sub-fields of affective computing for both speech and eye gaze modalities. Additionally, continuous affect prediction experiments using speech and eye gaze modalities are presented.  A baseline system is proposed using open source software, the performance of which is assessed on a publicly available audio-visual corpus. Further system performance is assessed in a cross-corpus and cross-lingual experiment. The experimental results suggest that eye gaze is an effective supportive modality for speech when used in a bimodal continuous affect prediction system. The addition of eye gaze to speech in a simple feature fusion framework yields a prediction improvement of 6.13\% for valence and 1.62\% for arousal.
\end{abstract}


\end{frontmatter}

\section{Introduction} 
Affective computing involves the computational analysis, recognition, prediction, and synthesis of emotion. This field  brings together research ranging from artificial intelligence to social science \citep{poria_review_2017}. Within affective computing, emotion recognition involves the computational recognition of an emotion that has been expressed by a subject. The emotions to be recognized may be classed, for example, as happy, sad, or angry. Multi-dimensional classes such as high/medium/low arousal or valence are also possible. Continuous affect prediction, another form of emotion computation, is the task of predicting a continuous numerical value for emotion dimensions, examples of which include arousal and valence. Arousal is a measure of how calming or exciting an experience is and valence is a measure of how positive or negative an experience is \citep{kossaifi_afew-va_nodate}. 

Affective computing tasks incorporating speech are now well developed. There are nearly thirty years of literature available on this topic \citep{fernandez_recognizing_2011}. A recent review carried out by \cite{feraru_cross-language_2015} showed that 66\% of all international languages are represented by an affective speech dataset. These speech datasets may be focused on emotion classification \citep{burkhardt_database_2005, soleymani_multimodal_2012} or continuous affect prediction within emotion dimensions such as arousal and valence \citep{mckeown_semaine_2012, ringeval_introducing_2013, valstar_avec_2014}. The emotions within the datasets may be acted, elicited, or natural as specified by the corpus data collection protocol. Within continuous affect prediction, speech has been found to perform well for arousal prediction, but less so for valence \citep{mencattini_continuous_2017, ringeval_prediction_2015}. Research into speech as a modality for affective computing is facilitated by the availability of research tools such as openSMILE \citep{eyben_recent_2013}, which is used to extract affective feature sets from speech. Examples of such feature sets include AVEC 2014 \citep{valstar_avec_2014}, ComParE  \citep{schuller_interspeech_2016}, and GeMAPS \citep{eyben_geneva_2016}.

Increasingly, multi-modal approaches to emotion classification and continuous affect prediction have been employed by the affective computing research community \citep{nicolaou_continuous_2011, ranganathan_multimodal_2016, ringeval_prediction_2015, thushara_multimodal_2016}. Datasets for the training and evaluation of multi-modal systems include YouTube \citep{morency_towards_2011}, SEMAINE \citep{mckeown_semaine_2012}, MAHNOB-HCI \citep{soleymani_multimodal_2012}, RECOLA \citep{ringeval_introducing_2013}, and AVEC 2014 \citep{valstar_avec_2014}. However, multi-modal affect recognition that includes eye gaze features extracted from video has not received much attention from the research community, based on available literature. This is surprising given the cost effective, non-intrusive nature of extracting data for this modality and the amount of audio-visual corpora available today. The AVEC 2016 \citep{valstar_avec_2016} challenge did provide OpenFace \citep{baltrusaitis_openface:_2016} eye gaze approximation features from video, but they were only provided for the depression corpus and not for the affect prediction challenge. Affect prediction using eye gaze from video deserves further investigation as shown by \cite{odwyer_continuous_2017}, particularly for continuous prediction of valence.

This work has two key contributions. Firstly it presents a review of research within the emotion classification and continuous affect prediction sub-fields of affective computing for both speech and eye gaze modalities. Feature sets, classification and regression methodologies, and performance levels in the literature are reviewed. When carrying out the literature review, a gap in the use of eye gaze for continuous affect prediction was observed. Secondly this work presents unimodal and bimodal continuous affect prediction experiments using speech and eye gaze from audio-visual sequences. Based on these experiments,  a bimodal speech and eye gaze affective computing system using the CURRENNT tool kit \citep{weninger_introducing_2015} is proposed. The results obtained clearly show the benefits of using eye gaze from video combined with speech in a bimodal continuous affect prediction system. Evaluations of the system include cross-corpus and cross-lingual experiments with promising results for arousal prediction.

This paper is structured as follows. Section 2 provides a review of affective computing using speech. In Section 3, multi-modal affective computing that includes speech information is reviewed. Section 4 presents a review of the observed gap in the literature, namely, affective computing using eye gaze. The proposal of a novel speech and eye gaze continuous affect prediction system, along with evaluation data, is presented in section 5. Section 6 presents the results and discussion of this work. Concluding remarks end this paper in Section 7.

\section{Unimodal affective computing using speech}
This section presents a review of emotion recognition and continuous affect prediction using speech. Data and methods used are presented, along with the results achieved for affective computing systems that were based only on speech features. A summary of each section of this review is given in Tables \ref{table:1} and \ref{table:2}.

\subsection{Emotion recognition using speech}
Emotion recognition involves the computational recognition of an emotion that has been expressed by a subject. The emotions to be recognized are classed discretely, such as happy, sad, or angry. Within the reviewed literature, both intra-corpus and cross-corpus emotion recognition experiments are presented. Intra-corpus evaluation is where an individual corpus is split into known training and unknown test partitions, the partitions are then used for machine learning model training and testing respectively. Cross-corpus evaluation involves multiple different corpora in varying configurations, for example, a model training partition might be gathered from one corpus and an unseen test set from a different corpus. This section reviews different methods applied to speech emotion recognition.

\cite{fayek_evaluating_2017} investigated the application of end-to-end deep learning for speech emotion recognition. The IEMOCAP \citep{busso2008iemocap} corpus was used for the experiments and feedforward neural network (FF), convolutional neural network (CNN), and long short-term memory recurrent neural network (LSTM-RNN) topologies were employed for evaluation and comparison. The results presented showed CNN performed best for the network architectures evaluated for the recognition of anger, happiness,
sadness, and neutral emotion classes. CNN was also shown to outperform deep neural network (DNN) and extreme learning machine (ELM), support vector machine (SVM), and hierarchical binary decision tree approaches for frame-based emotion recognition. A 60.89\% test set unweighted average recall (UAR) was achieved for frame-based emotion recognition using the CNN system.

\cite{motamed_speech_2017} introduced an optimized brain emotional learning model (BEL) that merged an adaptive neuro-fuzzy inference system (ANFIS) and multilayer perceptron (MLP) model for speech emotion recognition. The ANFIS was intended to model the human amygdala and orbitofrontal cortex in order to make rules that were passed to the MLP network. Mel-frequency cepstral coefficients (MFCCs) \citep{davis_and_mermelstein_1980} from speech were used as input to the system for the recognition tasks, which were performed on the Berlin EMO-DB \citep{burkhardt_database_2005}. The proposed algorithm performed better on average over all emotions when compared to ANFIS, MLP, BEL, BEL based on learning automata (BELBA), K-nearest-neighbour (KNN), and SVM approaches (72.5\% accuracy for anger, happiness, and sadness). KNN and SVM approaches used for comparison did outperform the proposed algorithm for both anger and happiness class recognition however.

\cite{c.k._hybrid_2017} presented work on higher order spectral features and feature selection approaches for emotion and stress recognition from speech. The authors added 28 bispectral and 22 bicoherence features to the Interspeech 2010 speech feature set \citep{schuller2010interspeech} and reported improved emotion recognition when compared to the Interspeech set alone. Additionally, the authors carried out biogeography-based optimisation (BBO), particle swarm optimisation (PSO), and BBO PSO hybrid optimisation techniques for feature selection. The feature additions and feature selection techniques were assessed using speaker-dependent, speaker-independent, male-dependent, and female-dependent approaches on the EMO-DB \citep{burkhardt_database_2005}, SAVEE \citep{haq2008audio}, and SUSAS \citep{hansen_getting_1997} corpora on an intra-corpus basis. SVM and ELM algorithms were used for model generation. The best result achieved on the EMO-DB \citep{burkhardt_database_2005} set resulted in improvements compared to previous work on that set for speaker independent experiments (93.25\% recognition rate). Also, a 100\% recognition rate was achieved using this method in a speaker-dependent experiment. These results were achieved using BBO optimisation and a SVM learning scheme.

\cite{chakraborty_knowledge-based_2016} evaluated a knowledge-based framework for the recognition of angry, happy, neutral, and sad emotion classes from speech. The proposed framework adds linguistic and time-lapse information about the conversation to the speech features. To check the validity of the approach the authors provided this linguistic and time-lapse information to annotators of an emotional corpus. It was observed that agreement between the annotators, as measured by Fleiss' Kappa \citep{l_fleiss_measuring_1971}, improved during the annotation process when the linguistic and time-lapse information was provided. This preliminary experiment was carried out on the Interactive Voice Response Speech Enabled Enquiry System (IVR-SERES) \citep{bhat_deploying_2013}. The utterance datasets used for the final experiments included one acted dataset, EMO-DB \citep{burkhardt_database_2005}, and two spontaneous emotion datasets, IVR-SERES \citep{bhat_deploying_2013} and Call Center \citep{Kopparapu_2014_NAC_2667132}. Within the experimental framework, the authors multiplied each emotion by a weight vector depending on how long the utterance was. The authors also extracted words from speech using automatic speech recognition (ASR) and then assigned a prominence measure, for example +, -, or 0, to words associated with specific emotions. The prominence measure contains information about how much emotion is contained in a word. The speech features for the system were that of the Interspeech 2009 Challenge \citep{schuller2009interspeech}. The performance of the system always increased with the addition of the time-lapse and emotion prominence from linguistic content. SVM, ANN, and KNN machine learning schemes were used for the experiments. The best results obtained were 82.1\% (IVR-SERES), 78.1\% (Call Center), and 87.3\% (Emo-DB) using a combination of all of the machine learning methods for classification.

\cite{vlasenko_tendencies_2016} investigated cross-corpus arousal classification using the VAM \citep{grimm_vera_2008} dataset for training and the EMO-DB \citep{burkhardt_database_2005} dataset for testing. During the experiments the authors trained models on data from the full VAM dataset, and two subsets of this dataset, VAM I, and VAM II. The VAM I and VAM II subsets contained very intense and intense emotions respectively. The authors describe the intensity in the VAM subsets as how well the emotions were conveyed. A hidden Markov model (HMM) approach was taken for low/high arousal classification. The authors concluded that there were big classification performance gaps for arousal models trained on spontaneous data with different emotional intensities. The entire VAM dataset for training provided the best test set performance UAR of 86\%.

\cite{song_cross-corpus_2016} investigated cross-corpus and cross-lingual emotion recognition using transfer learning combined with novel non-negative matrix factorization (NMF) on EMO-DB \citep{burkhardt_database_2005}, eNTERFACE \citep{martin_enterface_2006}, and FAU Aibo \citep{schuller2009interspeech} emotional speech corpora. EMO-DB and FAU Aibo corpora were the German datasets for the experiment and eNTERFACE represented English. Transfer graph regularized NMF (TGNMF) and transfer constrained NMF (TCNMF) additions to NMF were presented and evaluated for emotion recognition. The Interspeech 2010 feature set \citep{schuller2010interspeech} was the affective speech feature set used for the experiments. In addition to the proposed approaches, other algorithms that were used for comparison included transfer component analysis (TCA), conventional NMF, graph-regularized NMF (GNMF), and constrained NMF (CNMF). The proposed TGNMF and TCNMF methods always outperformed the other algorithms used for cross-corpus emotion recognition, with TCNMF performing best. TCNMF emotion recognition rates from the experiments ranged from 36.81\% (anger) to 74.81\% (disgust) on EMO-DB for a model trained using the eNTERFACE dataset.

\cite{dai_emotion_2015} proposed a support vector regression-based (SVR-based) method for emotion recognition on vocal social media in terms of position-arousal-dominance (PAD) emotion dimension estimation followed by categorical emotion mapping. There were 25 proposed emotional features gathered from speech for the experiments, which included prosody, spectral, and sound quality features. The authors then carried out model training using 180 chats from historical data on WeChat \footnote{\label{myfootnote1}https://web.wechat.com/}, a Chinese vocal social media platform. Following this the authors used the same WeChat group to create a test set for their model. The emotion recognition accuracy acheived was 82.43\% on average across happy, sad, angry, surprised, afraid, and neutral emotion classes. The reported PAD emotion dimension estimation error on this test set was 13.76\%. The authors then tested the developed model on a cross-corpus basis where a test set was gathered from QQ \footnote{\label{myfootnote2}https://im.qq.com/index.shtml}, another Chinese vocal social media platform. The aim of the cross-corpus testing was to assess model generalisability. The results from the cross-corpus test increased the model error by an absolute increase of 11.24\% above the initial WeChat error. The authors claimed that the personal features of the groups on social media had a significant impact on the estimation of PAD values from the model.

Modulation spectral features (MSF) were presented and their performance compared against short-time spectral features, MFCC and perceptual linear prediction (PLP), for the task of emotion recognition by \cite{wu_automatic_2011}. The authors also added the MSFs to prosodic features to obtain improvement in emotion recognition rates. In addition, the authors carried out a continuous affect prediction experiment where the system is reported to have achieved performance comparable to human annotators. The Berlin EMO-DB \citep{burkhardt_database_2005} and VAM \citep{grimm_vera_2008} datasets were used for the experiments. SVM and SVR machine learning schemes were used for classification and regression respectively. The results presented for the Berlin dataset indicate that MSF can outperform MFCC and PLP features for emotion classification tasks. Also, during a complimentary feature experiment, PLP, MFCC, and MSF were added to prosodic features for emotion recognition. The authors found that, on average, MSF outperformed PLP or MFCC when added to prosodic features over all emotion classes for emotion recognition. The authors also carried out a cross-database evaluation where a model was trained using the VAM dataset and then tested using the Berlin dataset. The proposed MSF feature set was used along with prosodic features. The recognition task for this experiment included continuous valence, arousal, and dominance level assessment and discrimination between two classes: anger versus joy, anger versus sadness, and anger versus fear. The results ranged from 58.6\% to 100\% with anger versus sadness being the best performing discrimination pair. 

\cite{fernandez_recognizing_2011} evaluated a model developed to infer affective categories using a hierarchical graphical model in the form of Dynamic Bayesian Networks (DBN) within their cross-corpus experiments. The hierarchical graphical model was intended to depict changes in prosodic-acoustic parameters over time (dynamic) and time-scales (hierarchical). The authors used 105 prosodic features gathered from speech, which included duration, intonational, loudness, and voice-quality features for model input. The authors gathered an acted corpora of emotions (afraid, angry, happy, neutral, and sad) which was used for model validation. CallHome \citep{canavan_callhome_1997}, and BT call center \citep{Durston01oasisnatural} corpora were employed for the model testing experiments. The model testing corpora contained natural, spontaneous emotion that was intended to be a more realistic assessment of the developed model. The emotion classes for model testing included high-arousal and negative-valence,  high-arousal and positive-valence,  low-arousal and negative-valence, and low-arousal and positive-valence. A neutral classification region was employed for some experiments. The best performance evaluation of the model achieved a 70\% emotion classification rate.

Cross-corpus emotion recognition using speech was investigated by \cite{schuller_cross-corpus_2010}. The motivation for this research was based on the view that an overestimation of machine learning performance was presented in the literature due to single, intra-corpus training and testing being employed. The corpora for the experiments included Danish Emotional Speech (DES) \citep{engbert_documentation_2007}, EMO-DB \citep{burkhardt_database_2005}, Speech under Simulated and Actual Stress (SUSAS) \citep{hansen_getting_1997}, eNTERFACE \citep{martin_enterface_2006}, Audio-visual Interest Corpus (AVIC) \citep{schuller_audiovisual_2007}, and smartKom \citep{steininger_development_2002}. The authors used SVM for the classification of up to six emotion classes. They also discriminated between positive or negative valence and high or low arousal within their experiments. Speaker, corpus, and speaker-corpus normalization techniques were used for the experiments with speaker normalization performing best. The highest performing test set result was achieved on the acted emotion dataset EMO-DB where 2-class emotion classification achieved  a median 70\% UAR. In general the results indicated that UAR performance dropped with a higher number of emotion classes. The authors concluded that cross corpus emotion recognition extends well only to acted data in clearly defined scenarios, for example controlled room acoustics, noise, and microphone-to-subject distance. The authors stated that corpus construction needs to improve to address the aforementioned issues and increase the generalisability of the emotion recognition models.

The review of speech emotion recognition showed that there are numerous corpora available for model research, validation, and testing \citep{haq2008audio, grimm_vera_2008, canavan_callhome_1997, Durston01oasisnatural, engbert_documentation_2007, burkhardt_database_2005, hansen_getting_1997, martin_enterface_2006, schuller_audiovisual_2007, steininger_development_2002}. Speech features used for input to emotion recognition models include MFCC \citep{motamed_speech_2017, wu_automatic_2011}, prosody \citep{dai_emotion_2015, wu_automatic_2011, fernandez_recognizing_2011}, and voice-quality \citep{fernandez_recognizing_2011, dai_emotion_2015} features. The use of a pre-compiled, publically available speech feature set was investigated as part of the work undertaken by \cite{c.k._hybrid_2017}, but this is not as common as manual feature extraction according to the reviewed literature. The novel approach taken taken by \cite{fayek_evaluating_2017} includes end-to-end deep learning that removes the need for speech feature extraction, lowering  the human effort required for emotional speech model building. Novel speech features from the literature worthy of further investigation include modulation spectral features \citep{wu_automatic_2011}, and bispectral and bicoherence features \citep{c.k._hybrid_2017}. While SVM machine learning approaches have been widely used for speech emotion recognition \citep{c.k._hybrid_2017, chakraborty_knowledge-based_2016, wu_automatic_2011, schuller_cross-corpus_2010}, there is a trend toward neural network modelling of emotional speech \citep{fayek_evaluating_2017, motamed_speech_2017}. A gap observed in the literature was that no author incorporated an explicit time-shift for ground-truth annotations at a frame level prior to model training and testing. The human annotation process adds a delay to the ground-truth annotations provided with corpora, and studies within contionuous affect prediction are incorporating these delays into ground-truth as shifts back in time \citep{valstar_avec_2016, he_multimodal_2015}. This is an area that merits further investigation for speech emotion recognition.

\begin{landscape}
\begin{table}[ht]
\scriptsize
\caption{Summary of methods for emotion recognition in speech} 
\centering 
\begin{tabular}{c c c c c c} 
\hline\hline 
Author(s) & Learning Scheme & Comment & Dataset(s) & Emotion Classes & Performance \\ [0.5ex] 
\hline 
\citep{fayek_evaluating_2017} & CNN & Frame-level recognition & IEMOCAP
& \begin{tabular}{c} Anger, happiness,\\sadness, neutral \end{tabular} & 60.89\% UAR \\ 
\citep{motamed_speech_2017} & NN & ANFIS addition to NN & EMO-DB
& \begin{tabular}{c} Anger, happiness,\\sadness \end{tabular} & 72.5\%   \\
\hline 
\citep{c.k._hybrid_2017} & SVM & BBO feature optimisation 
& \begin{tabular}{c} EMO-DB\\\\SAVEE\\\\SUSAS\end{tabular}
& \begin{tabular}{c} Anxiety, disgust,\\happiness, boredom,\\neutral, sadness, anger\\ 
Anxiety, disgust, fear\\neutral, sadness,\\surprise, happiness \\ 
Angry, lombard, loud,\\neutral \end{tabular}
& \begin{tabular}{c} 93.25\%\\\\62.38\%\\\\90.09\%\end{tabular} \\
\hline 
\citep{chakraborty_knowledge-based_2016} & SVM+ANN+KNN & Word emotion prominence added
& \begin{tabular}{c} EMO-DB \\ IVR-SERES \\ Call Center \end{tabular}
& \begin{tabular}{c} Anger, happy, sad,\\neutral\end{tabular}
& \begin{tabular}{c} 87.3\% \\ 82.1\% \\ 78.1\%\end{tabular} \\
\hline 
\citep{vlasenko_tendencies_2016} & HMM & Cross-corpus & EMO-DB & Low vs high arousal & 86\% UAR \\ 
\hline 
\citep{song_cross-corpus_2016} & NMF, transfer learning & Cross-corpus, cross-lingual 
& \begin{tabular}{c} EMO-DB\\\\FAU Aibo\\\\eNTERFACE \end{tabular}
& \begin{tabular}{c} \begin{tabular}{c}Anxiety, disgust,\\happiness, boredom,\\neutral, sadness, anger\\\end{tabular}  \\ Neutral, non-neutral \\ Happiness, anger,\\disgust, fear,\\sadness, surprise \end{tabular}
& \begin{tabular}{c} 52.10\%\\\\47.25\%\\\\44.86\% \end{tabular} \\
\hline 
\citep{dai_emotion_2015} & SVR+emotion mapping & Cross-corpus & QQ & HAPPY & 25\% error \\
\hline 
\citep{wu_automatic_2011}& SVM, SVR & Modulation spectral features & EMO-DB 
& \begin{tabular}{c} Anger vs joy  \\ Anger vs sadness \\ Anger vs fear \end{tabular}
& \begin{tabular}{c} 63.6\% \\ 100\% \\ 83.7\%  \end{tabular} \\
\hline 
\citep{fernandez_recognizing_2011} & DBN & Hierarchical graphical model 
& \begin{tabular}{c} Call Home \\ Call Center \end{tabular}
& \begin{tabular}{c} low/high arousal and\\positive/negative valence\\ high arousal-negative valence\\(binary 1/0) \end{tabular}
& \begin{tabular}{c} 44.1\% \\ 70\% detection rate,\\30\% false alarm  \end{tabular} \\
\hline 
\citep{schuller_cross-corpus_2010} & SVM & Cross-corpus, various normalization 
& \begin{tabular}{c} EMO-DB \\ DES \\ eNTERFACE \\ SMARTKOM \end{tabular}
& \begin{tabular}{c} 2 emotion classes \\ 3 emotion classes \\ 4 emotion classes \\ Arousal and valence levels \end{tabular}
& \begin{tabular}{c} 70\% UAR \\ 50\% UAR \\ 39\% UAR \\ 50\% UAR \end{tabular} \\
\hline 
\end{tabular}
\label{table:1} 
\end{table}
\end{landscape}

\subsection{Continuous affect prediction using speech}
\begin{figure}
  \includegraphics[width=\linewidth]{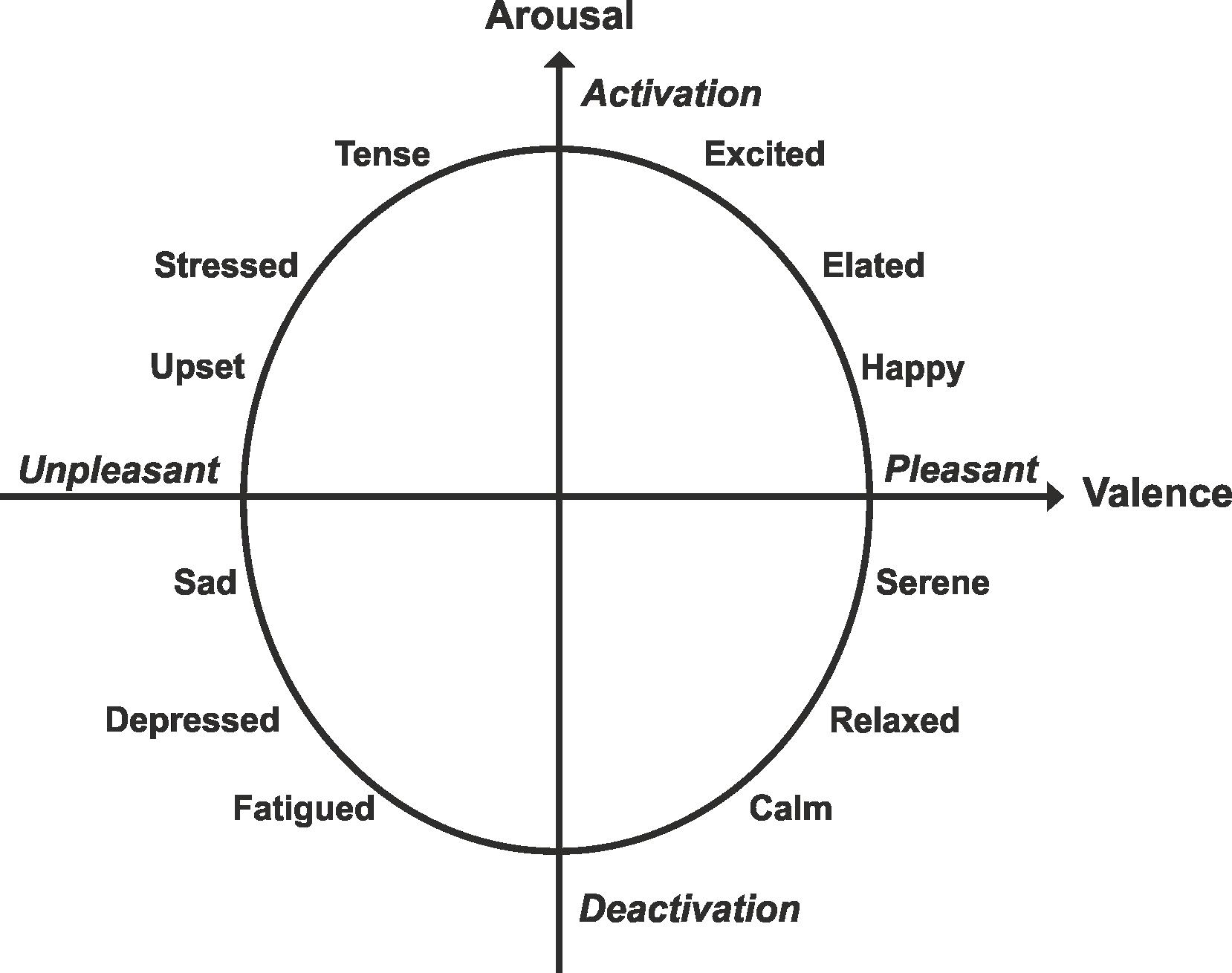}
  \caption{Arousal-Valence diagram from Abhang and Gawali, (2015)}
  \label{fig:1}
\end{figure}

Continuous affect prediction in speech is the task of predicting continuous numerical values for emotion dimensions. Numerical values for each emotion dimension typically range between +/- 1.0 for each speech frame. Emotion dimensions include arousal, which is a measure of how calming or exciting an experience is, and valence, which is a measure of how positive or negative an experience is \citep{kossaifi_afew-va_nodate}. Some common emotions are shown plotted on the arousal-valence plane in Figure \ref{fig:1}. Only intra-corpus experiments for continuous affect prediction have been found in the literature. The review of continuous affect prediction using speech is important for this work, as emotion dimensions, corpora, feature sets, and performance evaluations used inform the speech and eye gaze continuous affect prediction experiments presented in Section 5.

\cite{mencattini_continuous_2017} used single-speaker-regression-models (SSRMs) with the ComParE speech feature set \citep{schuller_interspeech_2016} for arousal and valence prediction on the RECOLA corpus \citep{ringeval_introducing_2013}. The regression techniques used for the experiment included partial least squares (PLS) and SVR. The authors incorporated weighted averaging of annotator ground-truth, annotator reaction lag, and feature selection for positive or negative arousal and positive or negative valence, which they called quadrant-based temporal division (QBTD). A concordance correlation coefficient (CCC) fusion measure was used to decide which SSRMs to include in the cooperative regression model (CRM) for the final test set predictions. The results showed that the SVR performed better than the PLS regression method for SSRM. However, the PLS method performed best for the CRM, which was the more general predictor. The authors suggest that SVR is more prone to over-fitting and simpler regression techniques such as PLS may be more suitable for machine learning techniques such as boosting, which is a similar ensemble method to the CRM method presented by the authors.

In \cite{gupta_predicting_2016}, the correlations between depression and affect were studied. Self-reported depression scores were incorporated into a continuous affect prediction system that used the AVEC 2014 \citep{valstar_avec_2014} database and speech features. Statistically significant correlations between affect dimensions and depression severity were found using a Student's t-test at the 5\% level. The authors based their subsequent investigation of an affect prediction system incorporating depression severity on these results. The final average results presented across arousal, valence, and dominance emotion dimensions were 0.33 correlation on the Freeform subset of AVEC 2014 and 0.52 on the Northwind subset. Including the depression severity score was found to improve the affect dimension correlations when compared with the authors' baseline system, which did not include the depression scores.

A bag-of-audio-words (BoAW) approach to continuous affect prediction of arousal and valence on the RECOLA \citep{ringeval_introducing_2013} dataset was presented in \cite{schmitt_at_2016}. The BoAW was created from MFCCs only and SVR was used for prediction from the codebook. The performance of BoAW was compared with that of functionals calculated from MFCCs and also against an end-to-end CNN approach to continuous affect prediction from the literature \citep{trigeorgis_adieu_2016}. The results showed the BoAW performed better than functionals on the test set. Additionally, the BoAW final arousal result in terms of CCC = 0.753, and valence = 0.465 was found to be better than a CNN approach from the literature \citep{trigeorgis_adieu_2016}.

\cite{zhang_facing_2016} carried out arousal and valence prediction on the RECOLA \citep{ringeval_introducing_2013} corpus using SVR on speech after non-stationary additive and convolutional noise was added to speech. The experiment used the ChiME15 \citep{barker_third_2015} database for additive noise and a convolution microphone input response (MIR) from a Google Nexus and room input response (RIR) for convolutional noises to be applied. The authors carried out feature enhancement on the noisy speech using LSTM-RNN and used this for comparison with the noisy speech input that they used as a baseline. SVR was used for affect prediction, temporal window sizes of 8 seconds, a temporal window step rate of 0.04 seconds, and a ground-truth delay of 4 seconds was applied to the speech input. The experiments included additive noise and smartphone noise continuous affect prediction performance. The smartphone noise consisted of MIR, RIR, or varying levels of ChiME noise. In addition two feature enhancement methods were employed using LSTM-RNN: (1) matched, in which several feature enhancement models were trained with different noise conditions, and (2) mixed, in which one feature enhancement model was trained with different noise conditions. The results showed that matched feature enhancement performed better than mixed feature enhancement. The CCC values for the matched condition were 0.596 for arousal and 0.223 for valence when ChiME15 noise was included, and 0.684 for arousal and 0.162 for valence when smartphone noise was included. The feature enhancement methods always improved affect recognition when compared with the baseline results.

\cite{asgari_automatic_2014} used speech, prosody, and text features as inputs to SVR systems for continuous affect prediction experiments on their own recorded dataset of children posing as actors in a short story re-enactment. Speech performed best for arousal prediction with a product moment correlation score of 0.86 and text performed best for valence with a correlation of 0.57. The authors concluded that the arousal of speech can be measured reliably, but not valence. It was also suggested that better text features need to be developed for valence dimension prediction.

The review of continuous affect prediction clearly shows arousal and valence as the emotion dimensions of choice for two-dimensional affect prediction \citep{mencattini_continuous_2017, ringeval_introducing_2013, asgari_automatic_2014, gupta_predicting_2016, schmitt_at_2016, zhang_facing_2016, valstar_avec_2014}. However, the AVEC 2014 corpus \citep{valstar_avec_2014} and the experiment presented by \citep{gupta_predicting_2016} included dominance as a third dimension, along with arousal and valence. The affective corpora for the experiments included AVEC 2014 \citep{valstar_avec_2014} and RECOLA \citep{ringeval_introducing_2013}. Pre-compiled speech feature sets for affect prediction from the literature included ComParE \citep{schuller_interspeech_2016} and the AVEC 2014 speech feature set \citep{valstar_avec_2014}. The majority of works reviewed carried out performance evaluation using CCC as a metric \citep{mencattini_continuous_2017, schmitt_at_2016, zhang_facing_2016} and this is justification for the use of CCC in this paper. No cross-corpus continuous affect prediction experiments could be found by the authors of this paper during the review. As such, continuous affect prediction needs cross-corpus experiments such as those undertaken in the speech emotion recognition community \citep{schuller_cross-corpus_2010, fernandez_recognizing_2011, dai_emotion_2015, song_cross-corpus_2016, vlasenko_tendencies_2016} to advance the field. 

\begin{table}[ht]
\scriptsize
\caption{Summary of continuous affect prediction methods for arousal (Ar), valence (Val), and dominance (Dom) prediction in speech} 
\centering 
\begin{adjustwidth}{-3.6cm}{}
\begin{tabular}{c c c c c} 
\hline\hline 
Author(s) & Learning Scheme & Comment & Dataset(s) & Performance \\ [0.5ex] 
\hline 
\citep{mencattini_continuous_2017} & PLS & Cooperative regression model, QBTD & RECOLA & CCC = 0.7 (Ar), 0.2 (Val) \\
\citep{gupta_predicting_2016} & SVR & Depression score inclusion
& \begin{tabular}{c} AVEC 2014 \\ (Northwind) \end{tabular} & COR = 0.52 (Avg. of Ar, Val, Dom) \\
\citep{schmitt_at_2016} & SVR & BoAW & RECOLA & CCC = 0.753 (Ar), 0.465 (Val) \\
\citep{zhang_facing_2016} & SVR & Noise addition, feature enhancement & RECOLA & CCC = 0.684 (Ar), 0.223 (Val) \\
\citep{asgari_automatic_2014} & SVR & Speech, prosody, text, ASR text & \citep{asgari_automatic_2014} & COR = 0.86 (Ar), 0.57 (Val) \\
\hline 
\end{tabular}
\end{adjustwidth}
\label{table:2} 
\end{table}

\section{Multi-modal affective computing using speech}
This section presents a review of multi-modal emotion recognition and continuous affect prediction using speech. Multi-modal affective computing is the study of more than one mode, or form of communication, during affective computing tasks. Of interest to this review are multi-modal systems that include speech. Again, the focus of this review within affective computing is emotion recognition, or recognizing an emotion displayed, and continuous affect prediction, which is the prediction of numerical values for emotion dimensions. The multi-modal systems reviewed include  video, context, and physiological features. The physiological features in the literature include EEG, eye gaze, EDA, heart rate, ECG, body movement, posture, or facial gesture features. Summaries of each section of the review are given in Tables \ref{table:3} and \ref{table:4}.

\subsection{Multi-modal emotion recognition using speech}
Emotion recognition involves the computational recognition of emotion. The literature review of multi-modal emotion recognition using speech is intended to highlight the features commonly used with speech in multi-modal systems, along with machine learning methods used and emotion recognition performances achieved.

\cite{nguyen_deep_2017} carried out audio-visual emotion recognition using 3-dimensional CNNs (C3Ns) and deep-belief networks (DBNs) on the eNTERFACE corpus \citep{martin_enterface_2006}. Spatio-temporal features were gathered using C3Ns from both audio and visual modalites and these features were passed further as input to audio and video DBNs. The audio and video DBNs were later fused using score-level fusion to produce the final emotion class decision. Input feature vectors for the C3Ns included a log-power spectra feature matrix with a 257 x 72 dimension for the audio C3N and a face region image extracted using a modified Viola-Jones algorithm from video frames for the video C3N. The modified Viola-Jones algorithm used by the authors employed a face cropping region the same as the previous frame if the face could not be detected. An emotion recognition rate of 82.83\% was achieved on average across anger, disgust, fear, happiness, sadness, surprise emotion classes.

\cite{poria_convolutional_2016} investigated multi-modal sentiment and emotion recognition from audio-visual and text sequences using temporal deep CNN. The IEMOCAP corpus \citep{busso2008iemocap} was used for the unimodal and multi-modal emotion recognition experiments presented. The temporal deep CNN proposed combined \textit{t} and \textit{t + 1} images, where \textit{t} is an image index in time, into one image feature vector and additionally used an RNN to model spatial and temporal dependencies for the CNN. A total of 6,373 speech features were extracted using openSMILE \citep{eyben_recent_2013} from audio for the experiments, and a positive or negative sentiment feature was extracted from text using a neural network approach. For multi-modal fusion, the authors used multiple kernel learning (MKL), proposed by \cite{subrahmanya_sparse_2010}. The authors describe multiple kernel learning as a fusion method whereby features are placed in groups, and then each group has its own kernel for model learning. From the experiments, the video emotion classifier performed best when compared against audio and text unimodal emotion classifiers. The multi-modal audio, text, and video classifier performed best overall when compared against the unimodal emotion recognition systems, however, the authors note that the multi-modal system performed only slightly better than the video classifier. The emotion recognition accuracies achieved for the multi-modal system were 79.2\% (angry), 72.22\% (happy), 75.63\% (sad), and 80.63\% (neutral).

\cite{yang_lightly-supervised_2016} performed utterance-level multi-modal emotion recognition on the USC CreativeIT database \citep{metallinou_usc_2016} using a semi-supervised learning approach. Pitch, energy, and MFCC features were gathered from audio, while hand gesture features were gathered from video for emotion classification. The audio-visual feature vector was combined using a proposed canonical correlation analysis (CCA) where both audio and video feature vectors were transformed using CCA projection vectors from each input modality prior to feature concatenation of each modality data. The authors used a multi-modal codebook, inspired by bag of words (BoW) approaches from text analysis, which used k-means clustering to create a codebook of multi-modal words for emotion recognition. The classification experiments included 2- and 3-class arousal and valence level discrimination. The arousal and valence classification results were 70.9\% for 2-class arousal and valence classification, and 53\% for 3-class arousal and valence classification using a code book of size 35. 

\cite{xie_new_2015} proposed a fusion method for multi-modal emotion recognition from audio-visual data. The experiments were carried out using prosody and MFCC features from audio, and Gabor filter and elastic body spline (EBS) facial features from video. The fusion method proposed incorporated both early and late fusion. For early fusion, separate audio and video modality feature transformation and fusion based on kernel entropy component analysis (KECA) \citep{jenssen_kernel_2010} was carried out. Following the early feature fusion, separate audio and video HMMs were used to classify the emotion. The individual HMM classifications were fused by way of score-level fusion to produce the final emotion decision. The score-level fusion proposed by \cite{xie_new_2015} is based on maximum correntropy criterion (MCC) optimisation, where correntropy is presented as ``a generalized similarity measure which has the stability to variation or noise". The MCC optimisation problem was solved using the algorithm presented by \cite{he_maximum_2011}. The emotion corpora used for the experiments included eNTERFACE \citep{martin_enterface_2006} and RTL \citep{wang_recognizing_2008}. Both of the corpora include anger, disgust, fear, sadness, surprise and happiness emotion ground-truth annotations. The results achieved over both corpora were 83\% average recognition accuracy over all emotions on the two corpora. Additional experiments showed that the bimodal system outperformed unimodal audio or video systems and that among the video features EBS features outperformed Gabor filter features.

\cite{soleymani_multimodal_2012} created an audio-visual affect database (MAHNOB-HCI) and investigated emotion recognition using speech, eye gaze, EEG, and physiological signals in their experiments. A Tobii X120 \footnote{\label{myfootnote4}https://www.tobii.com/} eye tracking device was used for eye gaze data capture from subjects. The authors divided arousal into classes of medium aroused, calm, and excited for the emotion recognition experiment and used SVM for classification. Valence was divided into classes of unpleasant, neutral valence, and pleasant. For multi-modal fusion, confidence measure summation
fusion was used. The emotion recognition results showed that eye gaze performed best during unimodal affect recognition experiments and a combination of eye gaze and EEG proved best overall. It must be stated however, that the speech modality may not have been maximally utilised as the subjects were required to watch emotion provoking video only. The unimodal eye gaze emotion classification accuracies achieved were 63.5\% for arousal and 68.8\% for valence. The EEG and eye gaze results were 67.7\% for arousal classification and 76.1\% for valence classification. While speech may not have been maximally used by \cite{soleymani_multimodal_2012}, it is of interest for the work presented by this paper that eye gaze performed best during the unimodal affect recognition experiments performed by \cite{soleymani_multimodal_2012}.

From the review of multi-modal emotion recognition using speech, it is clear that video is a popular modality for fusion with speech \citep{nguyen_deep_2017, poria_convolutional_2016, yang_lightly-supervised_2016, xie_new_2015, soleymani_multimodal_2012}. In addition, it is encouraging to see temporal consideration taken into account in the recent CNN approaches taken in the literature, by way of spatio-temporal feature gathering \citep{nguyen_deep_2017}, and temporal CNN and RNN combination \citep{poria_convolutional_2016}. 

The CCA fusion proposed by \cite{yang_lightly-supervised_2016} is interesting for bimodal affective computing research due to its simplicity. This fusion approach was shown to exploit the correlations between input modalities and further investigation of this technique is required using different machine learning techniques. A direct comparison of this work, and the fusion work undertaken by \cite{xie_new_2015}, against other approaches taken in the literature would be beneficial to advance multi-modal fusion research for affective computing.

Although not directly comparable, the average emotion recognition accuracy acheived by \cite{xie_new_2015} of 83\% is higher than that of \cite{nguyen_deep_2017} (82.83\%). While \cite{xie_new_2015} used more complex fusion methods, for example, both feature fusion by KECA and score-level fusion by way of MMC optimisation,  compared with \cite{nguyen_deep_2017} who simply used score-level fusion, \cite{nguyen_deep_2017} does use a more complex machine learning configuration of two C3Ns and two DBNs compared with two HMMs used by \cite{xie_new_2015}. However, the result acheived by \cite{xie_new_2015} is reported on both eNTERFACE \citep{martin_enterface_2006} and RTL \citep{wang_recognizing_2008} corpora, compared with only eNTERFACE for \cite{nguyen_deep_2017}. Reporting of the result acheived by \cite{xie_new_2015} on the eNTERFACE corpus only would yield a directly comparable result to that of \cite{nguyen_deep_2017}. This could offer interesting discussion on the complexity of fusion and learning methods, and their respective performance for multi-modal emotion recognition.

\begin{landscape}
\begin{table}[ht]
\scriptsize
\caption{Summary of multi-modal emotion recognition methods} 
\centering 
\begin{tabular}{c c c c c c} 
\hline\hline 
Author(s) & Learning Scheme & Comment & Dataset(s) & Emotion Classes & Performance \\ [0.5ex] 
\hline 
\citep{nguyen_deep_2017} & DBN & C3N spatio-temporal features used & eNTERFACE 
& \begin{tabular}{c} Anger, disgust, fear,\\happiness, sadness, surprise\end{tabular} & 82.83\%  \\
\citep{poria_convolutional_2016} & Temporal CNN/RNN & MKL fusion & IEMOCAP 
& \begin{tabular}{c} Angry\\Neutral\end{tabular}
& \begin{tabular}{c} 79.2\%\\80.63\%\end{tabular} \\
\citep{yang_lightly-supervised_2016} & BoW (k-means) & CCA fusion & USC CreativeIT & Arousal and valence levels & 70.9\% \\
\citep{xie_new_2015} & HMM & KECA early fusion, MCC late fusion &  eNTERFACE, RTL 
& \begin{tabular}{c} Anger, disgust, fear,\\happiness, sadness, surprise\end{tabular} & 83\%   \\
\citep{soleymani_multimodal_2012} & SVM & Cross-corpus & MAHNOB-HCI
& \begin{tabular}{c} Arousal level\\Valence level\end{tabular}
& \begin{tabular}{c} 67.7\%\\76.1\%\end{tabular} \\
\hline 
\end{tabular}
\label{table:3} 
\end{table}
\end{landscape}

\subsection{Multi-modal continuous affect prediction using speech}
Continuous affect prediction is the task of predicting continuous numerical values for emotion dimensions. There is an increasing interest by the affective computing community in multi-modal, continuous affect prediction as is evident by research challenges such as AVEC 2014 \citep{valstar_avec_2014}, AV+EC 2015 \citep{ringeval_av+ec_2015}, AVEC 2016 \citep{valstar_avec_2016}. Additionaly, there is now a large number of affective corpora for multi-modal research including AFEW-VA \citep{kossaifi_afew-va_nodate}, AVEC 2014 \citep{valstar_avec_2014}, RECOLA \citep{ringeval_introducing_2013}, and SEMAINE \citep{mckeown_semaine_2012}. Multi-modal continuous affective computing research using speech includes additional modalities in the overall system. Text, context, EEG, electro-cardiogram (ECG), electro-dermal activity (EDA), eye gaze, facial expression, gesture, posture, and video have all been used with speech for affect prediction. The variety provided for input data, advances in computer vision tools such as OpenFace \citep{baltrusaitis_openface:_2016}, and the fact that this is such an active area of research greatly motivates the work presented here in this paper.

\cite{stratou_multisense_2017} presented a framework designed to support multi-modal affective computing. The framework integrates open-source software, licensed software, and hardware support (camera, microphone, depth sensor) to achieve a unified framework for research and real-time applications for affective computing. Multi-modal input for the framework can include eye gaze, head pose, skeleton, speech, prosody, dialogue, and context. The framework incorporates context-based assessment at a local level by event creation and logging, such as when a question is asked, and at a global level by scenario definitions, for example, a job interview. The authors present a use case demonstrating the benefit of context-based features for automatic psychological distress analysis within a healthcare interview application. Subjects, which included US army veterans and general population, interacted with a virtual agent, SimSensei, and a behavioural report was produced after the user had finished their conversation with the virtual agent. A distress level prediction correlation of 0.7448 was achieved for a system that did not use context-based features. A distress level prediction correlation of 0.8822 was achieved for an automatic system that included context-based features, demonstrating the benefit of features calculated using context information.

\cite{brady_multi-modal_2016} were continuous affect prediction challenge winners for their work in the AVEC 2016 challenge \citep{valstar_avec_2016}. The authors used prosody, MFCC, and shifted delta cepstrum features from audio, in addition to the baseline audio features provided with the challenge, as input to SVR for arousal and valence prediction from audio. From the video modality, convolutional neural network (CNN) features were used as input to a recurrent neural network for arousal and valence regression. For prediction from the physiological channel, the authors opted to use the baseline features provided for ECG, skin conductance rate (SCR), and skin conductance level (SCL). LSTM regression features were trained from the baseline HRHRV and EDA features. The authors used Kalman filter fusion of audio, video, and physiological modalities for their multi-modal submission on the test set. The multi-modal fusion included model approximation, for example arousal or valence prediction, in addition to sensor channel measurements in the measurement matrix. The multi-modal results achieved on the test set were CCC values of 0.770 for arousal and 0.687 for valence.

Work presented in \cite{he_multimodal_2015} illustrates the deep BLSTM-RNN approach that these authors have taken in winning the AV+EC 2015 challenge \citep{ringeval_av+ec_2015}. The challenge was performed on the RECOLA dataset. Baseline features included with the dataset included speech, appearance and geometric video features, and physiological measures, which were comprised of ECG and EDA measurements. The authors added functionals of low-level descriptors extracted from speech using the YAFFE toolbox \citep{Mathieu2010YAAFEAE} to the speech baseline features and local phase quantization from three orthogonal planes (LPQ-TOP) from video to the video baseline features. The machine learning scheme used for prediction for each modality was BLSTM-RNN. Modality predictions were then fused by first applying Gaussian smoothing after which the prediction was passed to a final BLSTM-RNN for the final predictions. The CCC values achieved on the RECOLA test set were 0.747 for arousal and 0.609 for valence prediction.

A recent continuous affect prediction experiment undertaken by \cite{ringeval_prediction_2015} aimed to predict arousal and valence on the RECOLA \citep{ringeval_introducing_2013} corpus using neural networks. The modalities employed for this work included speech, video, ECG, and EDA. The machine learning schemes that were used included feed-forward neural networks, LSTM-RNN, and BLSTM-RNN. The CURRENNT tool kit \citep{weninger_introducing_2015} was used for network creation and training. BLSTM-RNN was the best performing network from the experiments. The authors experimented with various temporal window sizes for feature calculation, their findings suggesting that valence required a longer time window for optimum system performance. The CCC performance achieved was 0.804 for arousal and 0.528 for valence prediction.

Work presented by \cite{kachele_inferring_2014} demonstrated their approach to the AVEC 2014 continuous affect prediction and depression classification challenges \citep{valstar_avec_2014}. For the continuous affect prediction challenge, the authors investigated task dependent pattern templates for the prediction of each dimension. In addition to the challenge-provided features, the authors included application dependent meta knowledge features that included subject ID, gender, subject movement, subject age, estimated subject socio-economic status, and pixel statistics, among others. SVR and eigenvalue decomposition (EVD) with and without subject clustering approaches were used for the dimensional prediction from the audio-visual data. The subject clustering into 3 groups was carried out using a Ward’s distance measure based on the challenges depressive state features 1 - 29. The best Pearson correlations achieved were 0.6330 for arousal, and 0.5697 for dominance using an EVD and SVR approach with subject clustering. The best valence prediction result, a 0.5869 Pearson correlation was achieved using SVR with subject clustering. 

\cite{nicolaou_continuous_2011} carried out continuous arousal and valence prediction experiments on the SEMAINE \citep{mckeown_semaine_2012} corpus using speech, facial features, and shoulder gesture modality inputs. The SEMAINE \citep{mckeown_semaine_2012} corpus is a spontaneous emotion dataset in which audio-visual recordings of a subject interacting with a particular emotional character (happy, sad) have been recorded. The experiments included a comparison of machine learning algorithms and multi-modal fusion techniques. The authors also presented an output-associative fusion framework designed to exploit correlations and covariances between arousal and valence for model development. Experimental results showed BLSTM-RNN outperforming SVR during unimodal experiments. Additionally, speech was the best performing modality for arousal but the worst performing modality for valence within the unimodal investigation. For the final fusion experiments, output-associative fusion outperformed model and feature-level fusion of features. The highest reported correlation scores of 0.796 for arousal and 0.643 for valence were achieved using features from all modalities combined as input to a BLSTM-RNN utilising the proposed output-associative fusion framework. 

Multi-modal continuous affect prediction using speech is now both a well developed field, and an active area of research. Affect prediction performances acheived by \cite{nicolaou_continuous_2011}, \cite{ringeval_prediction_2015}, \cite{he_multimodal_2015}, and \cite{brady_multi-modal_2016} have achieved good results, given the challenging, non-acted emotion corpora used \citep{schuller_cross-corpus_2010}. BLSTM-RNN is widely used \citep{ringeval_prediction_2015, he_multimodal_2015, nicolaou_continuous_2011} for continuous affect prediction. In particular, \cite{nicolaou_continuous_2011} showed BLSTM-RNN outperforming SVR in his experiments and his findings are considered in the approach taken in the work presented in this paper. The work undertaken by \cite{kachele_inferring_2014} and \cite{stratou_multisense_2017} showed the importance of including context information during affect prediction. The result acheived by \cite{kachele_inferring_2014} is particularly impressive, as the Northwind subset of AVEC 2014 \citep{valstar_avec_2014} does not explicitly elicit an emotional response; this is in contrast to other common emotion corpora such as SEMAINE \citep{mckeown_semaine_2012}, RECOLA \citep{ringeval_introducing_2013}, and the AVEC 2014 \citep{valstar_avec_2014} Freeform subset, which have explicit emotion elicitation protocols. Recent work on affective computing software tools by \cite{stratou_multisense_2017} shows that eye gaze is an emerging modality for affective computing. Unfortunately, from the review undertaken, there were no cross-corpus or cross-lingual continuous affect prediction experiments available. This gap in the literature presents a new research opportunity for continuous affect prediction.

\begin{table}[ht]
\scriptsize
\caption{Summary of multi-modal continuous affect prediction methods for arousal (Ar), valence (Val), and dominance (Dom) prediction} 
\centering 
\begin{adjustwidth}{-4.1cm}{}
\begin{tabular}{c c c c c} 
\hline\hline 
Author(s) & Learning Scheme & Comment & Dataset(s) & Performance \\ [0.5ex] 
\hline 
\citep{stratou_multisense_2017} & Linear regression & Distress prediction, context features & \cite{stratou_multisense_2017} & COR = 0.8822 \\ 
\citep{brady_multi-modal_2016} & SVR, RNN & SDC features, Kalman filter fusion & RECOLA & CCC = 0.770 (Ar), 0.687 (Val) \\ 
\citep{he_multimodal_2015} & BLSTM-RNN & Gaussian smoothed model fusion & RECOLA & CCC = 0.747 (Ar), 0.609 (Val) \\
\citep{ringeval_prediction_2015} & BLSTM-RNN & Fusion, varying temporal windows & RECOLA & CCC = 0.804 (Ar), 0.528 (Val) \\
\citep{kachele_inferring_2014} & SVR & Application meta-knowledge & AVEC 2014 & COR = 0.5965 (Avg. Ar, Val, Dom) \\
\citep{nicolaou_continuous_2011} & BLSTM-RNN & Output-associative fusion & RECOLA & CCC = 0.796 (Ar), 0.643 (Val) \\
\hline 
\end{tabular}
\end{adjustwidth}
\label{table:4} 
\end{table}

\section{Affective computing using eye gaze}
From the review of the literature on emotion recognition and continuous affect prediction using speech, a gap  has been observed concerning the use of speech combined with eye gaze in multi-modal affective computing systems. A further review is presented here for affective computing with eye gaze as an input modality. This section concludes with suggestions for combining speech and eye gaze as input modalities for affective computing. The eyes are a rich source of sociological \citep{itier_neural_2009}, neuropsychological \citep{lappi_eye_2016}, and arousal-level \citep{partala_pupil_2003} information. Eye-based features applied to affective computing include: eye blink \citep{soleymani_multimodal_2012, lanata_eye_2011}, eye gaze \citep{odwyer_continuous_2017, soleymani_multimodal_2012, lanata_eye_2011}, visual focus of attention \citep{zhao_facial_2011}, pupillometry \citep{soleymani_multimodal_2012}, and pupil size variation \citep{lanata_eye_2011}. Other measurements, including eye saccade (saccadometry), are also possible to gather from the eyes for affective computing research.

\subsection{Emotion recognition using eye gaze}
Unimodal emotion recognition using eye gaze was investigated by \cite{aracena_neural_2015} using an EyeLink 1000 \footnote{\label{myfootnote3}http://www.sr-research.com/eyelinkII.html} eye tracking headset. Eye gaze data were gathered from the eye tracker with emotion classification carried out using neural networks. The best performing method from this experiment included a decision tree with neural networks. The decision tree neural network correctly recognized emotion classes as either positive, negative or neutral with an average accuracy of 53.6\% for 4 male subjects on a subject-independent basis. Subject-dependent scores for the decision tree neural network ranged from 62.3\% to 78.4\% classification accuracy.

\cite{zhao_facial_2011} cited psychological research suggesting a correlation of direct eye gaze with angry and happy emotions in human-to-human communication. The authors also claimed that sadness and fear are associated with averted gaze. A geometrical eye and nostril model was used to identify averted gaze and direct gaze in video input. This eye gaze data was added to facial expression to improve emotion classification in angry, sad, fear, and disgust recognition. Happy and surprise recognition was not improved when eye gaze was added to facial features for the classification. The authors mention that facial illumination of subjects was controlled in the experiment, which limits the applicability of the results to more natural environments. 

\cite{lanata_eye_2011} assessed whether eye gaze tracking and pupil size variation could provide useful cues to discriminate between emotional states induced in subjects viewing images at different arousal content levels. The emotional states were defined as neutral and high arousal. Subjects were provided images from the international affective picture system (IAPS) \citep{article} which was intended to evoke these responses. A new wearable and wireless eye gaze tracker called HATCAM was proposed. Recurrence quantification analysis (RQA) \citep{webber2005recurrence}, along with fixation time and pupil area detection features were used for classification using k-nearest-neighbours (KNN). The experimental results showed that the proposed hardware and features could be used to discriminate between users experiencing different emotion stimuli. Emotion recognition rates of 90\% for neutral images and approximately 80\% for high arousal images were achieved.

\subsection{Continuous affect prediction using eye gaze}
\cite{odwyer_continuous_2017} investigated unimodal continuous affect prediction in terms of arousal and valence using eye gaze. The Freeform subset of the AVEC 2014 dataset \citep{valstar_avec_2014} was used for the experiments along with the speech features provided with AVEC 2014 for baseline comparison. An SVR learning scheme was employed for both speech and eye gaze experiments and a new eye gaze feature set for continuous affect prediction was proposed. The results showed that eye gaze could perform well for valence prediction when compared to speech, but speech was the far better predictor of arousal. The Pearson's correlation scores from the experiment were: eye gaze valence 0.3318, speech valence 0.3107, eye gaze arousal  0.3322, speech arousal 0.5225.

\subsection{Affective computing using speech and eye gaze}
From the review of the literature, it appears that continuous affect prediction using the combination of speech and eye gaze data from audio-visual sources is still largely unexplored. The inclusion of speech in multi-modal continuous affective prediction systems is common, and good results have been achieved in the literature \citep{ringeval_prediction_2015, brady_multi-modal_2016}. Combining speech and eye gaze, or more generally, including eye gaze in affective computing systems is receiving greater interest from the research community \citep{stratou_multisense_2017, odwyer_continuous_2017, aracena_neural_2015, zhao_facial_2011, lanata_eye_2011}. Recent projects such as OpenFace \citep{baltrusaitis_openface:_2016} and MultiSense \citep{stratou_multisense_2017} now make eye gaze features from video easily accessible to multi-modal affect systems research. However, the literature has focused on emotion recognition using eye gaze, only one paper could be identified as investigating eye gaze for continuous affect prediction \citep{odwyer_continuous_2017}. Furthermore, the combination of speech and eye gaze in a bimodal continuous affect prediction system has not been explored. 

The next section details a proposal for a novel speech and eye gaze continuous affect prediction system based on LSTM-RNNs. The proposed system contains optimised ground-truth annotation delays to account for human perception of audio-visual sequences. The system training and performance evaluation data are also presented and discussed.

\section{A bimodal speech and eye gaze affect prediction system}
In this section, a bimodal speech and eye gaze continuous affect prediction system based on BLSTM-RNN and LSTM-RNN is proposed. Training and performance evaluation data are discussed.

\subsection{System framework}
\begin{figure}
  \includegraphics[width=\linewidth]{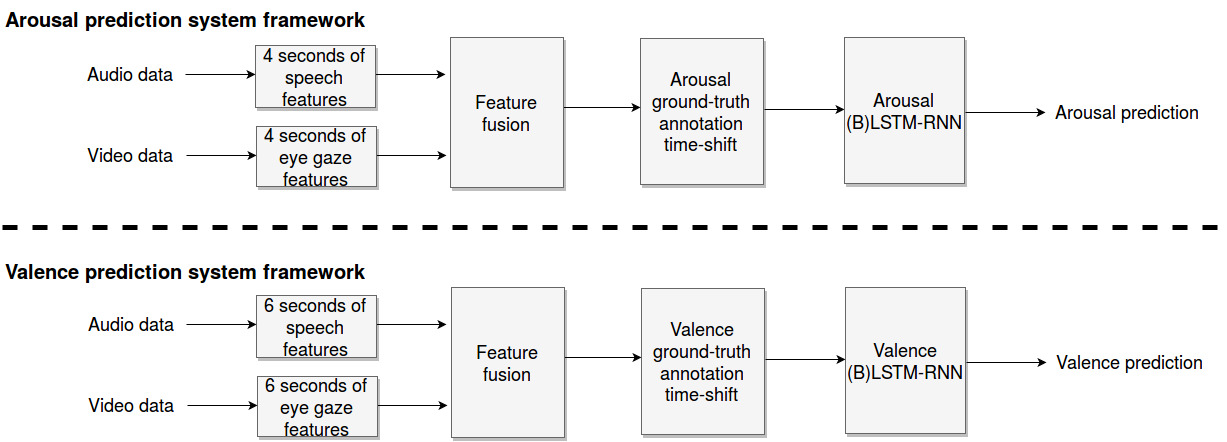}
  \caption{Proposed bimodal speech and eye gaze affect prediction system}
  \label{fig:2}
\end{figure}

The BLSTM and LSTM ((B)LSTM) neural networks presented are trained on a single-task basis. That is, any network is only trained to predict arousal or valence. The fusion framework used for the system is feature fusion, also known as early feature fusion, which is the concatenation of speech and eye gaze features into one large feature vector for each training or testing example. Additionally, the system framework is designed to achieve the best possible advantage from the ground-truth annotations for arousal and valence provided with each corpus, which the annotators provided in response to observed audio-visual recordings of subjects. Therefore, the system employs a ground-truth time-shift, back in time, prior to neural network training or testing. The optimal ground-truth time-shift is selected for the bimodal system, for both arousal and valence affect prediction tasks. A block diagram of the system framework is given in Figure \ref{fig:2}.

\subsection{Experimental approach}
\begin{figure}
  \includegraphics[width=\linewidth]{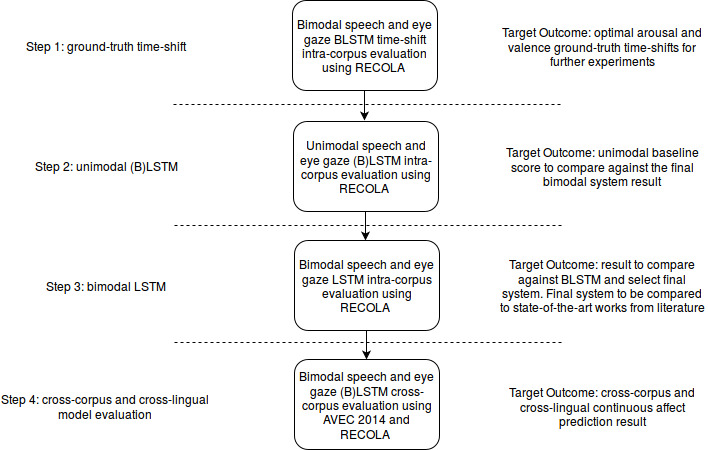}
  \caption{Summary of experimental steps and target outcomes in the design of the proposed speech and eye gaze continuous affect prediction system}
  \label{fig:3}
\end{figure}

Following from \cite{ringeval_prediction_2015}, single-task BLSTM networks are trained using the CURRENNT toolkit \citep{weninger_introducing_2015} for arousal and valence prediction. There are two hidden layers, the first with 40 nodes and the second with 30 nodes. Additionally, unidirectional LSTM-RNN models consisting of two hidden layers with 80 and 60 nodes respectively are generated for performance comparison with BLSTM and system selection. Bimodal speech and eye gaze (B)LSTM models are first created using a number of different ground-truth time-shifts in order to select the optimal shift to be applied to arousal and valence annotations for training the networks. The ground-truth time-shift experiment is followed by unimodal (B)LSTM speech and eye gaze model creation to evaluate the contribution of eye gaze to speech in the proposed affect prediction system. Finally, cross-corpus and cross-lingual continuous affect prediction is carried out using the best performing models from the uni- and bimodal intra-corpus experiments. BLSTM-RNN has the advantage of both past and future temporal context during network training. The unidirectional LSTM recurrent neural network variant only has the benefit of past context, or memory, as it is trained. The LSTM addition to RNN was first presented by \cite{hochreiter_long_1997} to avoid the vanishing gradient problem that RNNs suffer from. These neural networks contain what \cite{hochreiter_long_1997} presented as memory units that allow nodes to store context information during network training. Today, LSTM-RNN is widely used for affective computing model generation, with good performance being achieved in recent works such as \citep{ringeval_prediction_2015}, \citep{he_multimodal_2015}, \citep{valstar_avec_2016}, and \citep{brady_multi-modal_2016}. The experimental steps and their target outcomes are summarised in Figure \ref{fig:3}.

\subsection{Audio-visual corpora}
The RECOLA corpus contains audio, visual, and physiological recordings of spontaneous affectively coloured dyad interactions in French. The subjects within each dyad were required to perform a task together requiring collaboration while being recorded. Human-provided numerical ground-truth annotations for arousal and valence are provided for each frame of each recording within the corpus for automatic prediction system training and performance evaluation.

The AVEC 2014 corpus \citep{valstar_avec_2014} is an audio-visual corpus recorded in German with ground-truth values provided  for arousal and valence for each frame of each recording in the corpus. In particular, the Freeform partition of this corpus is used for the cross-corpus experiment. The Freeform partition of this dataset has subjects interacting with a computer to answer emotionally provocative questions while audio-visual recordings were taken. The tasks, communication, emotion elicitation, and emotional responses for both AVEC and RECOLA corpora can be compared in Table \ref{table:5}. 

\begin{table}[ht]
\scriptsize
\caption{RECOLA and AVEC Freeform emotion corpora protocols} 
\centering 
\begin{adjustwidth}{-4cm}{}
\begin{tabular}{c c c c c} 
\hline\hline 
Corpus & Task & Communication & Emotion elicitation & Emotional response \\ [0.5ex] 
\hline 
AVEC 2014 \citep{valstar_avec_2014} & Responding to emotionally provocative questions & Human-computer & Explicitly provoked &  Elicited \\ 
RECOLA \citep{ringeval_introducing_2013} & Collaborative dyadic problem solving & Human-human* & Interaction and task response &  Spontaneous \\
[1ex] 
\hline 
*Used remote computer interaction
\end{tabular}
\end{adjustwidth}
\label{table:5} 
\end{table}

\subsection{Affective feature extraction}
Affective feature extraction is the process of calculating features from raw data provided from the input modalities of choice for a system. Features can include low-level descriptors, for example, energy or spectral parameters for speech \citep{eyben_geneva_2016}, and eye gaze distance \citep{odwyer_continuous_2017} for eye gaze. Statistics of the low-level descriptors are often calculated as part of the final feature set to be passed as input to machine learning models for classification or regression. The eGeMAPS \citep{eyben_geneva_2016} speech feature set was used for affective speech feature input in this work. Speech features are extracted using openSMILE \citep{eyben_recent_2013} with a calculation window of 4 seconds worth of frames for arousal dimension feature calculation and 6 seconds for valence dimension feature calculation. The feature calculation window sizes are the same as those used for the audio information channel in \citep{valstar_avec_2016}. The feature calculation windows were moved forward at a rate of 1 frame as was the case for \citep{valstar_avec_2016}. The eGeMAPS \citep{eyben_geneva_2016} feature set was used to provide baseline features in the AVEC 2016 Challenge \citep{valstar_avec_2016} and provides a total of 88 affective features from speech. For eye gaze affective features, the feature set presented by \cite{odwyer_continuous_2017}, containing a total of 31 features, is extracted. The affective eye gaze features from this set are listed in Table \ref{table:6}. Raw eye gaze data are gathered from video using OpenFace \citep{baltrusaitis_openface:_2016} and this is followed by feature extraction from the raw data using the same time windows and window rates as for the speech features. In summary, for the bimodal affect prediction experiments a total of 119 features are extracted for model training, validation, and testing.

\begin{table}[ht]
\scriptsize
\caption{Eye gaze feature set for continuous affect prediction from \cite{odwyer_continuous_2017}} 
\centering 
\begin{adjustwidth}{-1.6cm}{}
\begin{tabular}{c c} 
\hline\hline 
Data & Features \\ [0.5ex] 
\hline 
Eye gaze distance & 
\begin{tabular}{c} eye gaze approach ratio, \\ average eye gaze approach time in milliseconds \end{tabular} \\
\hline
Eye gaze scan paths & 
\begin{tabular}{c} average scan path length, \\ standard deviation of scan path lengths \end{tabular} \\
\hline 
Vertical and horizontal eye gaze coordinates & 
\begin{tabular}{c} average, inter quartile range 1-2, inter quartile range 2-3, \\ standard deviation, skewness, power spectral densities at \\
frequencies [0.011, 0.022, 0.033-0.044, 0.055-0.066, 0.077-0.133]Hz, \\ average of standard deviation of coordinates in each fixation zone, \\ standard deviation of standard deviation of coordinates in each \\ fixation zone
\end{tabular} \\
\hline 
Eye closure & 
\begin{tabular}{c} average eye close frame count, \\ standard deviation of eye close frame
count, \\ skewness of eye close frame count \end{tabular} \\ [1ex] 
\hline 
\end{tabular}
\end{adjustwidth}
\label{table:6} 
\end{table}

\subsection{Feature fusion}
\begin{figure}
  \centering
  \includegraphics[width=250px]{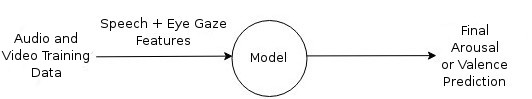}
  \caption{Speech and eye gaze feature fusion}
  \label{fig:4}
\end{figure}

\begin{figure}
  \centering
  \includegraphics[width=250px]{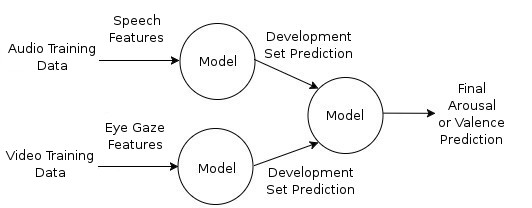}
  \caption{Speech and eye gaze decision fusion}
  \label{fig:5}
\end{figure}

Feature fusion is the process of combining features gathered from different modalities with the intention of improving the performance of machine learning systems. Feature fusion can be performed early, late, or both, with regard to the machine learning process. Early feature fusion, in the form of feature fusion, is simply the row-wise concatenation of features from each modality into one large feature vector prior to model training and testing. An illustration of feature fusion is given in Figure \ref{fig:4}. An example of late feature fusion is decision fusion. Decision fusion, shown in Figure \ref{fig:5}, is where multiple models are trained, one for each modality, after which the model decisions that are made on a portion of the training data, or development set, are combined to make a further model for the final test set classification or regression. The simplest fusion approach taken in \citep{ringeval_prediction_2015}, feature fusion, is employed for the work presented here. The feature fusion method was selected based on experimentation carried out by \cite{odwyer_eye_gaze_and_speech_2017}, where feature fusion was shown to perform best for the continuous prediction of arousal.

\subsection{Neural network training for affect prediction}
The (B)LSTM-RNN neural network training methodology follows that of \cite{ringeval_prediction_2015}. As such, inputs and ground-truth targets are normalised to zero mean and unit variance prior to neural network training. In addition, Gaussian noise with a standard deviation 0.1 is added to all input features prior to training. The neural network training takes place for a maximum of 100 epochs, and training stops if no improvement of the performance as measured by sum of squared errors is observed on the validation set for more than 20 epochs. The network learning rate and random seed hyperparameters are selected based on the lowest validation set error achieved during experimentation.

\begin{figure}
  \includegraphics[width=\linewidth]{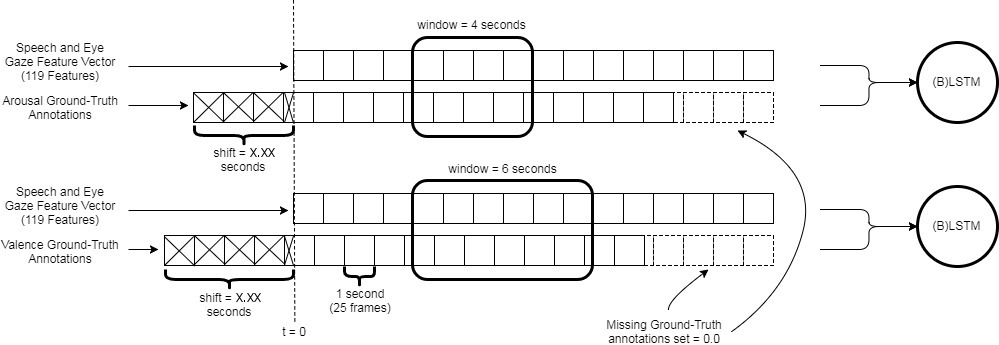}
  \caption{Arousal and valence ground-truth annotation time delay}
  \label{fig:6}
\end{figure}

One important difference between the proposed system training and that of \cite{ringeval_prediction_2015} is the time-shift of ground-truth annotations prior to network training. To take into account the delay in human reaction times when producing ground-truth annotations for the corpora, a shift back in time is applied to the ground-truth prior to concatenation with the speech and eye gaze features for network training and testing. \cite{ringeval_prediction_2015} argued that LSTM-RNN can encode temporal context during network training, which allows this machine learning method to overcome the need for ground-truth time-shift to account for human annotation lag. However, based on results obtained by \cite{he_multimodal_2015}, incorporating ground-truth delays prior to (B)LSTM network training, is used in the work presented in this paper. Therefore, the ground-truth annotation values for arousal and valence are shifted back in time for the start of each recording. At the end of each recording, missing annotation values are padded as 0.0 rated arousal and valence values. The annotation delay values for the experiments vary in range from +/-1 second around the audio modality ground-truth time-shifts proposed by \cite{he_multimodal_2015} for arousal and valence, respectively. The same delays are incorporated for both speech and eye gaze modalities given the early feature fusion framework being used. This process is illustrated in Figure \ref{fig:6}.

\subsection{Performance evaluation}
Performance evaluation of the proposed system on the RECOLA corpus \citep{ringeval_introducing_2013} is carried out using CCC \citep{lin_concordance_1989}. CCC combines a penalty of mean-squared error with a Pearson correlation as in Equation ~\eqref{CCC-def}, where \textit{x} represents the machine predicted values,  \textit{y} represents the ground-truth values, $\sigma$ is the covariance, $\sigma^2$ is the variance, and $\mu$ is the mean. The CCC gives a measure of agreement between the machine predicted values and the ground-truth values.

\begin{equation}
  \label{CCC-def}
  CCC = \frac{2 \sigma_{xy}}{\sigma^2_{x} + \sigma^2_{y} + (\mu_{x} - \mu_{y})^2} \
\end{equation}

In addition to intra-corpus evaluation for the bimodal system that is trained, validated and tested on the French language RECOLA corpus \citep{ringeval_introducing_2013}, a cross-corpus and cross-lingual evaluation is carried out. The AVEC 2014 corpus \citep{valstar_avec_2014} is used for this evaluation. The Freeform partition of this German language corpus is used to assess the performance of the bimodal network model that has been trained exclusively using the French language RECOLA corpus \citep{ringeval_introducing_2013}. An additional cross-corpus and cross-lingual evaluation of a system trained using AVEC 2014 and tested using RECOLA is also carried out.

\section{Results and discussion} 
In this section results and discussion are presented for bimodal (B)LSTM ground-truth time-shift, intra-corpus unimodal (B)LSTM and bimodal (B)LSTM experiments. The proposed bimodal system is compared against other work from the literature. Finally, results for the proposed system during the cross-corpus and cross-lingual experiment are presented and discussed.

\subsection{Ground-truth time-shift evaluations}
\begin{figure}
  \includegraphics[width=\linewidth]{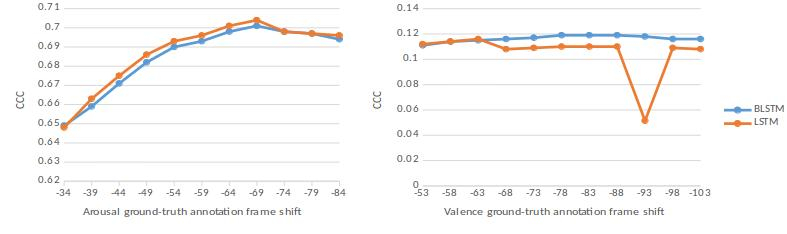}
  \caption{BLSTM-RNN arousal and valence prediction results using different ground-truth annotation time-shifts}
  \label{fig:7}
\end{figure}

For the ground-truth time-shift experiment, the ground-truth time shifts proposed by \cite{he_multimodal_2015} for the audio modality are used as a guide for the ground-truth time-shifts tested in this work. \cite{he_multimodal_2015} found a shift back in time of 59 frames for arousal and 78 frames for valence to be optimal when considering audio in a unimodal fashion on the RECOLA corpus \citep{ringeval_introducing_2013}. For the experiments conducted here, bimodal (B)LSTM-RNNs were trained on the training partition of the RECOLA corpus \citep{ringeval_introducing_2013}, and tested on the validation partition. Figure \ref{fig:7} shows the validation set performances of the arousal and valence (B)LSTM-RNNs with different time-shifts applied to the ground-truth annotations prior to training and testing.

Figure 7 clearly shows that altering the ground-truth annotation time-shift does have an effect on (B)LSTM-RNN performance. The highest performance achieved by both BLSTM and LSTM networks for arousal prediction had a 69 frame negative-shift applied to ground-truth annotations prior to training and testing. The highest performance achieved by BLSTM and LSTM networks for valence prediction had frame negative-shifts of 78 and 63 applied respectively. The RECOLA corpus \citep{ringeval_introducing_2013} was recorded at 25 frames per second, this results in a 2.76 seconds ground-truth time-shift for the arousal networks, and 3.12 (BLSTM) and 2.52 (LSTM) seconds ground-truth time-shift for the valence networks. These ground-truth time-shift values were used for the experiments described in the following sections.

\subsection{Intra-corpus evaluations}
Unimodal and bimodal LSTM and BLSTM network arousal and valence prediction performances achieved on the RECOLA corpus \citep{ringeval_introducing_2013} are shown in Table \ref{table:7}. The results for both LSTM and BLSTM systems clearly demonstrate the benefit of including eye gaze features with those of speech in a feature fusion framework for both arousal and valence prediction. The feature fusion BLSTM arousal prediction CCC of 0.754 is an increase in performance of 1.62\% when compared to the highest performing unimodal speech system. For valence, the feature fusion BLSTM CCC of 0.277 represents a 6.13\% performance increase when compared to the highest performing unimodal speech system. LSTM provided better performance than BLSTM for arousal and valence prediction using speech alone. This suggests that past memory is more relevant for neural network training when using unimodal speech input. The highest performing networks for arousal and valence prediction overall, the BLSTM networks, show that future context in addition to past context is important for neural network training for continuous affect prediction in a bimodal system. 

\begin{table}[ht]
\scriptsize
\caption{RECOLA intra-corpus emotion dimension prediction CCC and arousal (Ar) and valence (Val) neural network hyperparameters for unimodal and bimodal experiments} 
\centering 
\begin{tabular}{c c c c c c} 
\hline\hline 
Arousal & Valence & Learning Method & Modality & Random Seed & Learning Rate \\ [0.5ex] 
\hline 
0.732 & 0.261 & LSTM & Speech & \begin{tabular}{c} 1787452436 \\ 123456789 \end{tabular} & \begin{tabular}{c} $8x10^-5$ (Ar) \\ $5x10^-6$ (Val) \end{tabular} \\
0.232 & 0.02 & LSTM & Eye gaze & \begin{tabular}{c} 1787452436 \\ 123456789 \end{tabular} & \begin{tabular}{c} $7x10^-5$ (Ar) \\ $2x10^-6$ (Val) \end{tabular} \\
0.742 & 0.213 & LSTM & Feature Fusion & \begin{tabular}{c} 1787452436 \\ 1787452436 \end{tabular} & \begin{tabular}{c} $9x10^-6$ (Ar) \\ $3x10^-6$ (Val) \end{tabular} \\
0.735 & 0.255 & BLSTM & Speech & \begin{tabular}{c} 1787452436 \\ 123456789 \end{tabular} & \begin{tabular}{c} $1x10^-5$ (Ar) \\ $9x10^-7$ (Val) \end{tabular} \\
0.268 & -0.001 & BLSTM & Eye Gaze & \begin{tabular}{c} 1787452436 \\ 1787452436 \end{tabular} & \begin{tabular}{c} $3x10^-5$ (Ar) \\ $4x10^-6$ (Val) \end{tabular} \\
\textbf{0.754} & \textbf{0.277} & BLSTM & Feature Fusion & \begin{tabular}{c} 1787452436 \\ 1787452436 \end{tabular} & \begin{tabular}{c} $1x10^-5$ (Ar) \\ $1x10^-5$ (Val) \end{tabular} \\[1ex] 
\hline 
\end{tabular}
\label{table:7} 
\end{table}

\subsection{Comparison with previous approaches from the literature}
The (B)LSTM-based system proposed in this work is compared in Table \ref{table:8} with previous approaches to intra-corpus evaluation on the RECOLA \citep{ringeval_introducing_2013} corpus. The central contribution of the work presented in this paper is to demonstrate the benefit that eye gaze can have when combined with speech in a continuous affect prediction system. However, a comparison against other works puts the achieved results in context within the literature. Therefore, the comparison includes other multimodal results from the literature that used the GeMAPS \citep{eyben_geneva_2016} feature set for affect prediction on the RECOLA corpus \citep{ringeval_introducing_2013}. Table \ref{table:8} shows that the proposed bimodal speech and eye gaze affect prediction system outperforms \cite{he_multimodal_2015} for arousal prediction. Compared to \cite{brady_multi-modal_2016} however, the proposed bimodal system presented here does not provide a performance increase for arousal prediction. Of note for this work when compared against \cite{brady_multi-modal_2016} is that a much simpler fusion approach and a smaller feature vector was used. Valence prediction performance for the proposed system does need improvement to be comparable to the literature. However, the central focus of this study is what the eye gaze modality can add to speech for continuous affect prediction, which was clearly demonstrated in Table \ref{table:7}. 

\begin{table}[ht]
\scriptsize
\caption{Emotion dimension prediction CCC of proposed bimodal system compared with approaches from the literature} 
\centering 
\begin{adjustwidth}{-0.75cm}{}
\begin{tabular}{c c c c c} 
\hline\hline 
Arousal & Valence & Feature Count & Fusion Method & Authors \\ [0.5ex] 
\hline 
0.747 & 0.609 & 270 (arousal), 259 (valence) & Gaussian smoothed model fusion & \citep{he_multimodal_2015} \\ 
\textbf{0.770} & \textbf{0.687} & 622 & Kalman filter & \citep{brady_multi-modal_2016} \\
0.754 & 0.277 & 119 & Feature fusion & This work \\ [1ex] 
\hline 
\end{tabular}
\end{adjustwidth}
\label{table:8} 
\end{table}

\subsection{Cross-corpus and cross-lingual evaluations}
During cross-corpus and cross-lingual experiments, the highest performing neural network models trained and tested using the French RECOLA corpus \citep{ringeval_introducing_2013} were tested using the German AVEC 2014 Freeform corpus \citep{valstar_avec_2014}. As such, the feature fusion BLSTM-RNNs are selected for this task. Also, to complete the cross-comparisons, BLSTM arousal and valence networks were trained using AVEC 2014 Freeform \citep{valstar_avec_2014} and tested using RECOLA \citep{ringeval_introducing_2013}. The ground-truth time-shift had to be altered to 84 frames for arousal and 96 frames for valence for the AVEC 2014 corpus training, validation and test sets for the experiment due to it being recorded at 30 frames per second. The cross-corpus test set results from the best performing neural networks from intra-corpus validation set experiments are shown in Table \ref{table:9}.  

\begin{table}[ht]
\scriptsize
\caption{Proposed BLSTM system cross-corpus and cross-lingual emotion dimension prediction CCC and arousal (Ar) and valence (Val) neural network hyperparameters} 
\centering 
\begin{tabular}{c c c c c c} 
\hline\hline 
Arousal & Valence & Training corpus & Testing corpus & Random Seed & Learning Rate \\ [0.5ex] 
\hline 
0.263 & 0.097 & RECOLA & AVEC Freeform & \begin{tabular}{c} 1787452436 \\ 1787452436 \end{tabular} & \begin{tabular}{c} $1x10^-5$ (Ar) \\ $1x10^-5$ (Val) \end{tabular} \\
0.363 & -0.081 & AVEC Freeform & RECOLA & \begin{tabular}{c} 1787452436 \\ 3926481610 \end{tabular} & \begin{tabular}{c} $1x10^-5$ (Ar) \\ $1x10^-5$ (Val) \end{tabular} \\ [1ex] 
\hline 
\end{tabular}
\label{table:9} 
\end{table}

The cross-corpus and cross-lingual experiment undertaken provides a significant challenge for the neural network models developed. In contrast to \cite{schuller_cross-corpus_2010}, where Germanic languages only were used for cross-corpus emotion classification, this work uses Germanic (German) and Romance (French) languages for continuous affect prediction performance assessment. Additionally, the protocols used to compile the emotion corpora are also different, as presented in Table \ref{table:5}. The contribution of these results is that they are the first set of cross-corpus and cross-lingual results for continuous affect prediction from the literature reviewed. From the results in Table \ref{table:9}, clearly, arousal is more easily predicted across corpora and languages. Interestingly, networks trained using the German AVEC corpus \citep{valstar_avec_2014} and tested using the French RECOLA corpus \citep{ringeval_introducing_2013} performed 38.02\% better on average during arousal prediction than networks trained using the French RECOLA corpus and tested using the German AVEC corpus. It is recognised by the affective computing research community that valence is much more difficult to predict compared to arousal. The difficulty for valence prediction here is compounded by the cross-corpus and cross-lingual prediction task.

\section{Conclusions}
From the review of the literature on affective computing using speech, it is clear that considering eye gaze with speech is relatively unexplored for affective computing, specifically for continuous affect prediction of emotion dimensions. This paper proposed a new bimodal affect prediction system using speech and eye gaze input to a BLSTM learning scheme. The proposed system design is simple, based on openly available software including OpenFace \citep{baltrusaitis_openface:_2016}, openSMILE \citep{eyben_recent_2013}, and CURRENNT \citep{weninger_introducing_2015}, and does not require the use of expensive or intrusive hardware.

From the experiments presented, it is clear that eye gaze features extracted from video improves arousal and valence prediction system performance when combined with speech features. The results from the proposed system are promising for arousal dimension prediction with the BLSTM variant outperforming LSTM for bimodal affect prediction. The best performing network for valence prediction needs further development for the proposed framework to produce results comparable to the literature for this emotion dimension. Altering temporal window sizes for feature calculation across different modalities is one option to improve valence dimension prediction, based on work by \cite{ringeval_prediction_2015}, and \cite{valstar_avec_2016}. More complex modality fusion and feature selection methods, and ground-truth reliability weighting \citep{grimm_evaluation_2005} are other potential avenues of investigation to improve system performance for this difficult to predict \citep{ringeval_influence_2013} emotion dimension.

The novel cross-corpus results provided in this work serve as a baseline for researchers attempting cross-corpus, cross-lingual, and cross-protocol continuous affect prediction. The results achieved indicate that predicting arousal under these conditions is reasonable, but additional work is needed for the continuous affect prediction of valence during cross-corpus and cross-lingual experiments. From the results presented, it is hoped that the research community will consider eye gaze combined with speech in future multi-modal systems. Further evaluations including the proposed modalities within cross-corpus, cross-lingual, cross-task, and even cross-culture affective computing experiments are hoped for, to increase model generalisability and impact.

Future work will include ground-truth reliability weighting, a model fusion comparison against the proposed system, and transfer learning during cross-corpus experiments. These studies will allow investigation of the contribution of eye gaze when combined with speech for continuous affect prediction under further experimental conditions. Further cross-corpus continuous affect prediction studies on an intra-lingual basis and the addition of further non-intrusive modalities to the proposed (B)LSTM-RNN based system are also planned.

\section*{Acknowledgement}
Funding: This work was supported by the Irish Research Council [grant number GOIPG/2016/1572].

\section*{References}

\bibliography{mybibfile}
\break
\includegraphics[width=60px, height=70px]{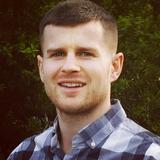}
Jonny O' Dwyer received the BEng degree in Software Engineering from Athlone Institute of Technology. He is currently a postgraduate MSc research student at Athlone Institute of Technology. His research interests include affective computing and machine learning. He is a Government of Ireland Postgraduate Scholar.

\includegraphics[width=60px, height=70px]{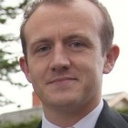}
Niall Murray is a faculty member of Engineering \& Informatics in Athlone Institute of Technology. He received the BE degree in Electronic Engineering from National University of Ireland, Galway, the MEng degree from University of Limerick and the PhD degree from AIT. He is the founder of and the principal investigator in the truly Immersive and Interactive Multimedia Experiences research group. He is a Science Foundation Ireland funded investigator in the Confirm Centre for Smart manufacturing and an associate investigator on the Enterprise Ireland-funded COMAND technology gateway. His research interests include immersive and multi-sensory multimedia communication, and quality of experience.

\includegraphics[width=60px, height=70px]{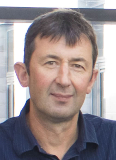}
Ronan Flynn received the BE degree in Electronic Engineering from University College Dublin, the MEng degree in Computer Systems from University of Limerick and the PhD degree from National University of Ireland, Galway. Dr. Flynn is currently a lecturer and researcher in the Faculty of Engineering \& Informatics in Athlone Institute of Technology. He has been involved in the evaluation of H2020 proposals submitted to the European Commision. He was on the management committee for European COST Action IC1206 "De-identification for privacy protection in multimedia content". His research interests include speech recognition, emotion recognition in speech and multi-modal affective computing.

\end{document}